\documentclass[aps,pra,10pt,tightenlines,longbibliography,notitlepage,superscriptaddress,amsmath,twocolumn,floatfix,eqsecnum]{revtex4-1}
\usepackage[dvipsnames]{xcolor}
\usepackage[colorlinks=true,bookmarks=false,linkcolor=NavyBlue,urlcolor=NavyBlue,citecolor=NavyBlue,breaklinks]{hyperref}
\usepackage{amsmath}
\usepackage{amsfonts}
\usepackage{graphicx}
\usepackage{mathtools}

\newcommand{\real}{\operatorname{Re}}

\newcommand{\intall}{\int_{-\infty}^{\infty}}
\newcommand{\ket}[1]{|#1\rangle}

\newcommand{\bra}[1]{\langle#1|}

\newcommand{\bk}[1]{\left(#1\right)}
\newcommand{\Bk}[1]{\left[#1\right]}
\newcommand{\BK}[1]{\left\{#1\right\}}

\newcommand{\trace}{\operatorname{tr}}

\newcommand{\expect}{\mathbb E}

\begin{document}

\title{Spectrum analysis with quantum dynamical systems}

\author{Shilin Ng}

\affiliation{Department of Physics, National University of Singapore,
  2 Science Drive 3, Singapore 117551}

\author{Shan Zheng Ang}

\affiliation{Department of Electrical and Computer Engineering,
  National University of Singapore, 4 Engineering Drive 3, Singapore
  117583}

\author{Trevor A.~Wheatley}
\affiliation{School of Engineering and Information Technology, University of New South Wales Canberra, ACT 2600, Australia}
\affiliation{Centre for Quantum Computation and Communication Technology, Australian Research Council}

\author{Hidehiro Yonezawa}
\affiliation{School of Engineering and Information Technology, University of New South Wales Canberra, ACT 2600, Australia}
\affiliation{Centre for Quantum Computation and Communication Technology, Australian Research Council}

\author{Akira Furusawa}
\affiliation{Department of Applied Physics, School of Engineering, The University of Tokyo, 7-3-1 Hongo, Bunkyo-ku, Tokyo 113-8656, Japan}

\author{Elanor H.~Huntington}
\affiliation{Centre for Quantum Computation and Communication Technology, Australian Research Council}
\affiliation{Research School of Engineering, College of Engineering and Computer Science, Australian National University, North Road, Acton, Canberra ACT 2600, Australia}

\author{Mankei Tsang}

\email{mankei@nus.edu.sg}
\affiliation{Department of Electrical and Computer Engineering,
  National University of Singapore, 4 Engineering Drive 3, Singapore
  117583}

\affiliation{Department of Physics, National University of Singapore,
  2 Science Drive 3, Singapore 117551}



\date{\today}


\begin{abstract}
  Measuring the power spectral density of a stochastic process, such
  as a stochastic force or magnetic field, is a fundamental task in
  many sensing applications. Quantum noise is becoming a major
  limiting factor to such a task in future technology, especially in
  optomechanics for temperature, stochastic gravitational wave, and
  decoherence measurements. Motivated by this concern, here we prove a
  measurement-independent quantum limit to the accuracy of estimating
  the spectrum parameters of a classical stochastic process coupled to
  a quantum dynamical system. We demonstrate our results by analyzing
  the data from a continuous optical phase estimation experiment and
  showing that the experimental performance with homodyne detection is
  close to the quantum limit. We further propose a spectral photon
  counting method that can attain quantum-optimal performance for weak
  modulation and a coherent-state input, with an error scaling
  superior to that of homodyne detection at low signal-to-noise
  ratios.
\end{abstract}

\maketitle
\section{Introduction}
Recent technological advances, especially in optomechanics
\cite{aspelmeyer14}, suggest that quantum noise will soon be the major
limiting factor in many metrological applications
\cite{braginsky}. Many tasks in optomechanics force sensing, including
thermometry, estimation of stochastic gravitational-wave background
\cite{ligo09}, and testing spontaneous wavefunction collapse
\cite{nimmrichter,diosi15}, involve the spectrum analysis of a
stochastic force, and the effect of quantum noise on such tasks has
been of recent interest \cite{nimmrichter,diosi15}. To study the
quantitative effect of experimental design on estimation accuracy, it
is important to use a rigorous statistical inference framework to
investigate the parameter estimation error. While there exist many
theoretical studies of quantum parameter estimation for thermometry
(see, for example,
Refs.~\cite{stace,marzolino,jarzyna,correa,nair_tsang}), their
application to more complex dynamical systems with broadband
measurements such as optomechanics remains unclear.

In this work, we propose a theoretical framework of spectrum-parameter
estimation with quantum dynamical systems, proving fundamental limits
and investigating measurement and data analysis techniques that
approach the limits. An outstanding feature of our work is the simple
analytic results in terms of basic power spectral densities (PSDs) in
the problem, such that they can be readily applied to optics and
optomechanics experiments. To illustrate our theory, we analyze a
recent experiment of continuous optical phase estimation and
demonstrate that the experimental performance using homodyne detection
is close to our quantum limit. We further propose a spectral photon
counting method that can beat homodyne detection and attain
quantum-optimal performance for weak modulation and a coherent-state
input. The advantage is especially significant when the
signal-to-noise ratio (SNR) is low, thus demonstrating the importance
of quantum-optimal measurements and coherent optical information
processing in the low-SNR regime for gravitational-wave astronomy
\cite{ligo16} and optical sensing in general.

\section{Quantum metrology}
\subsection{Parameter estimation}
Consider a quantum dynamical system with Hamiltonian $\hat H[X,t]$ as
a functional of a c-number hidden stochastic process $X(t)$, such as a
classical force.  Assume that the prior probability measure of $X(t)$
depends on a vector of unknown parameters $\theta$. Let $Y$ be the
quantum measurement outcome and $\check\theta(Y)$ be an estimator of
$\theta$ using $Y$.  The central error figure of interest is the
mean-square estimation error matrix, defined as
\begin{align}
\Sigma_{\mu\nu}(\theta) \equiv \expect_Y
\BK{\Bk{\check\theta_\mu(Y)-\theta_\mu}\Bk{\check\theta_\nu(Y)-\theta_\nu}},
\end{align}
with $\expect_Y$ denoting the expectation over the random variable
$Y$. Our goal here is to compute analytic results concerning $\Sigma$
and discover quantum measurement techniques that can accurately
estimate $\theta$.

For any unbiased estimator ($\expect_Y(\hat\theta) = \theta$), the
multiparameter Cram\'er-Rao bound states that
\begin{align}
\Sigma \ge j^{-1}(P_Y),
\label{crb}
\end{align}
where $j(P_Y)$ is the classical Fisher information matrix with respect
to the observation probability measure $P_Y$ \cite{vantrees}. The
matrix inequality means that $\Sigma - j^{-1}$ is
positive-semidefinite, that is,
$u_\mu(\Sigma-j^{-1})_{\mu\nu} u_\nu \ge 0$ for any real vector $u$
(Einstein summation is assumed throughout this paper).  For a quantum
system, let $\hat\rho(\theta)$ be a $\theta$-dependent density
operator and $\hat E(y)$ be the positive operator-valued measure
(POVM) that models the measurement, such that
\begin{align}
P_Y(y|\theta) = \trace\Bk{\hat E(y)\hat\rho(\theta)},
\end{align}
with $\trace$ being the operator trace. For dynamical systems,
$\hat\rho(\theta)$ can be obtained using the principles of
purification and deferred measurements
\cite{nielsen,twc,tsang_nair,tsang_open}. For the purpose of
spectrum-parameter estimation, we model $\hat\rho$ as
\begin{align}
\hat\rho(\theta) &= \expect_{X|\theta}
\BK{\hat U[X,T]\ket\psi\bra\psi \hat U^\dagger[X,T]},
\label{ensemble}
\end{align}
where
\begin{align}
\hat U[X,T]=\mathcal T \exp\BK{-\frac{i}{\hbar}\int_0^T dt \hat H[X,t]}
\end{align}
is the unitary time-ordered exponential of $\hat H$ with total
evolution time $T$, $\ket\psi$ is the initial quantum state, and the
expectation is with respect to the hidden process $X(t)$, the prior
probability measure of which depends on $\theta$. $\theta$ is called
hyperparameters in this context \cite{granade12}. For any POVM, a
quantum Cram\'er-Rao bound states that
\begin{align}
  j(P_Y) \le J(\hat \rho),
\label{qcrb}
\end{align}
where $J(\hat\rho)$ is the quantum Fisher information matrix with
respect to the symmetric logarithmic derivatives of $\hat\rho$
\cite{helstrom,holevo11,hayashi}.

\subsection{Extended convexity}
While quantum parameter estimation bounds for dynamical systems have
been studied previously in the context of low-dimensional systems such
as qubits (see, for example,
Refs.~\cite{brunelli11,benedetti14,gammelmark14}), $J$ is much more
difficult to evaluate analytically for multimode high-dimensional
dynamical systems under continuous measurements. To proceed, we
exploit a recently discovered property of $J$ known as the extended
convexity \cite{alipour}, which states that
\begin{align}
J(\hat\rho) \le \mathcal J\BK{\hat\sigma,P_Z}
\equiv 
\expect_{Z|\theta}\Bk{J\bk{\hat\sigma}} + j(P_Z),
\label{extended}
\end{align}
where $\{\hat\sigma,P_Z\}$ is any ensemble of $\hat\rho$ with elements
$\hat\sigma$ and mixing probability measure $P_Z$ such that
$\hat\rho(\theta) = \expect_{Z|\theta}[\hat\sigma(Z|\theta)]$.

The proof of extended convexity $J\le\mathcal J$ for one parameter in
Ref.~\cite{alipour} relies on the assumption that there exists an
optimal POVM attaining $j = J$. Such an assumption is questionable
however \cite{barndorff}, and here we use instead the strong concavity
of Uhlmann fidelity \cite{nielsen} to prove Eq.~(\ref{extended}) for
multiple parameters.  Let $\{\hat\sigma,P_Z\}$ be an ensemble for
$\hat\rho(\theta)$ such that
\begin{align}
\hat\rho(\theta) = \int dz P_Z(z|\theta)\hat\sigma(z|\theta).
\end{align}
Define the Uhlmann fidelity as
\begin{align}
F[\hat\rho,\hat\rho'] \equiv
\trace\sqrt{\sqrt{\hat\rho}\hat\rho'\sqrt{\hat\rho}}.
\end{align}
 The strong
concavity states that \cite{nielsen}
\begin{align}
F\Bk{\hat\rho(\theta),\hat\rho(\theta')} &\ge 
\int dz \sqrt{P_Z(z|\theta)P_Z(z|\theta')} 
\nonumber\\&\quad\times
F[\hat\sigma(z|\theta),\hat\sigma(z|\theta')].
\label{strong_concavity}
\end{align}
To relate $F$ to $J$, we use the fact \cite{hayashi}
\begin{align}
  F[\hat\rho(\theta),\hat\rho(\theta+\epsilon u)] = 1-\frac{\epsilon^2}{8}
  u_\mu J_{\mu\nu}(\hat\rho) u_\nu + o(\epsilon^2),
\label{fidelity_qfi}
\end{align}
where $\epsilon$ is a scalar, $u$ is any real vector with the same
dimension as $\theta$, and $o(\epsilon^2)$ denotes terms
asymptotically smaller than $\epsilon^2$. It is also known that
\cite{kailath1967}
\begin{align}
\int dz \sqrt{P_Z(z|\theta)P_Z(z|\theta+\epsilon u)} &= 
1-\frac{\epsilon^2}{8} u_\mu j_{\mu\nu}(P_Z) u_\nu 
\nonumber\\&\quad
+ o(\epsilon^2).
\label{bhatta_cfi}
\end{align}
Expanding $F[\hat\rho(\theta),\hat\rho(\theta')]$ and
$F[\hat\sigma(z|\theta),\hat\sigma(z|\theta')]$ in
Eq.~(\ref{strong_concavity}) using Eq.~(\ref{fidelity_qfi}), applying
Eq.~(\ref{bhatta_cfi}) to the right-hand side of
Eq.~(\ref{strong_concavity}), and comparing the $\epsilon^2$ terms on
both sides, we obtain
\begin{align}
u_\mu J_{\mu\nu}(\hat\rho) u_\nu \le u_{\mu} 
\BK{\expect_{Z|\theta}\Bk{J_{\mu\nu}(\hat\sigma)} + j_{\mu\nu}(P_Z)}u_\nu.
\end{align}
Since this holds for any $u$, we obtain the matrix inequality in
Eq.~(\ref{extended}). The classical simulation technique proposed in
Ref.~\cite{demkowicz} can be regarded as a special case of extended
convexity when $J(\hat\sigma) = 0$.

\subsection{\label{sec_dyn}Dynamical systems}
\begin{figure}[htbp!]
\centerline{\includegraphics[width=0.45\textwidth]{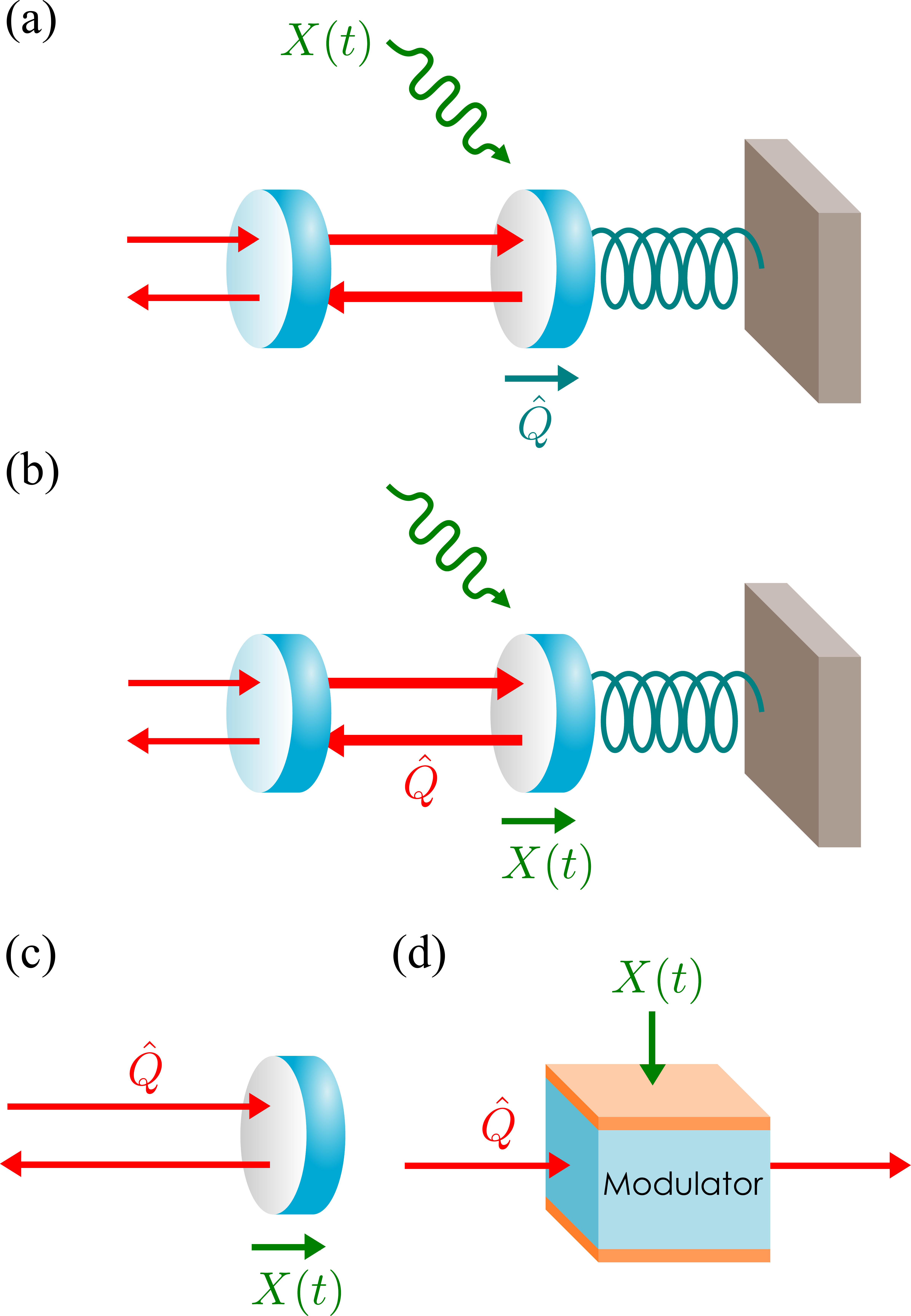}}
\caption{(Color online). Some examples of the hidden stochastic
  process $X(t)$ and generator $\hat Q$. (a) $X(t)$ is the classical
  force and $\hat Q$ is the mechanical position, (b) $X(t)$ is the
  c-number forced displacement and $\hat Q$ is proportional to the
  photon-number operator, (c) and (d) $X(t)$ is the phase modulation
  and $\hat Q$ is proportional to the photon-flux operator.}
\label{generator_process_examples}
\end{figure}

To compute simple analytic results for dynamical systems, we make
further assumptions.  Assume that $X(t)$ is zero-mean, Gaussian, and
stationary, with a PSD given by
\begin{align}
S_X(\omega|\theta) \equiv \intall d\tau
\expect_{X|\theta}[X(t)X(t+\tau)] \exp(i\omega \tau).
\end{align}
For the quantum system, we assume that the Hamiltonian is of the form
\begin{align}
\hat H = \hat H_0 - \hat Q X(t),
\label{H}
\end{align}
where $\hat Q$ is the quantum generator and $\hat H_0$ is the rest of
the Hamiltonian. For example, $X(t)$ can be the classical force on a
mechanical oscillator and $\hat Q$ can be the quantum position
operator, as depicted in Fig.~\ref{generator_process_examples}(a).

A modified purification technique can
transform the problem in the interaction picture and produce an
alternative and possibly tighter bound in terms of the optical
statistics alone \cite{tsang_open}.
For an optomechanical system, the Hamiltonian is of the form
\cite{aspelmeyer14}
\begin{align}
\hat H_{\rm OM} &= \hat H_{\rm M} + \hat H_{\rm O} + \hat h,
\end{align}
where $\hat H_{\rm M}$ is the mechanical Hamiltonian, $\hat H_{\rm O}$ is the
optical Hamiltonian, and $\hat h$ is the optomechanical
interaction Hamiltonian.  For example, if the mechanical oscillator
with position operator $\hat q$ interacts with one cavity optical mode
with photon-number operator $\hat n$,
$\hat h = -\hbar g_0 \hat n \hat q$, where $g_0$ is a coupling
constant. A classical force $f(t)$ on the mechanical oscillator leads
to a term $-\hat q f(t)$ in $\hat H_{\rm M}$, and if we assume $\hat U$ to
be the time-ordered exponential of $\hat H_{\rm OM}$, $f(t)$ can be
regarded as the hidden process and $\hat q$ the generator.

In practice, measurements are made on the optics and not the mechanics
directly, so one is free to modify the purification \cite{escher} by
applying any mechanical unitary to the optomechanical one
\cite{tsang_open}. To be specific, let $\hat U_{\rm OM}$ be the
time-ordered exponential of $\hat H_{\rm OM}$ and $\hat U_{\rm M}$ be
the time-ordered exponential of $\hat H_{\rm M}$. Since the POVM is
not applied to the mechanics, $\hat U\ket\psi\bra\psi \hat U^\dagger$
with $\hat U = \hat U_{\rm M}^\dagger \hat U_{\rm OM}$ is also a valid
purification for a given force \cite{tsang_open}.  $\hat U$ becomes
the time-ordered exponential of the interaction-picture Hamiltonian
\begin{align}
\hat H(t) &= \hat H_{\rm O} + \hat h_{\rm M}(t),
&
\hat h_{\rm M}(t) &\equiv \hat U_{\rm M}^\dagger(t)\hat h \hat U_{\rm M}(t).
\end{align}
For cavity optomechanics,
$\hat h_{\rm M}(t) = -\hbar g_0 \hat n \hat q_{\rm M}(t)$, where
$\hat q_{\rm M}(t)$ is the interaction-picture mechanical
position. For a linear mechanical system,
$\hat q_{\rm M}(t) = \hat q_{0}(t) + X(t)$, where $\hat q_{0}(t)$ is
the operator-valued homogeneous component as a function of the initial
position and momentum operators and $X(t)$ is the c-number
inhomogeneous component of the displacement due to the classical
force. We can hence take $X(t)$ to be the hidden process and
$\hat Q = \hbar g_0 \hat n$ to be the generator, obtaining uncertainty
relations between the displacement errors and the photon-number
fluctuations, as depicted in Fig.~\ref{generator_process_examples}(b).

In general, this interaction-picture purification method can be
applied to any linear system with Hamiltonian of the form
$\hat H_0 - \hat Q X(t)$, where $\hat Q$ is a canonical coordinate
operator and $\hat H_0$ is quadratic with respect to canonical
coordinates, as the effect of $X(t)$ remains a displacement operation
in any interaction picture.

Figure~\ref{generator_process_examples}(c) and (d) depict two other
examples of Eq.~(\ref{H}) in the context of optical phase modulation,
in which case $X(t)$ is the phase modulation on the optical beam and
$\hat Q$ is proportional to the photon-flux operator. Other examples
include the magnetometer, where $X(t)$ is an external magnetic field
and $\hat Q$ is a spin operator \cite{hall09}, and the voltmeter,
where $X(t)$ is an applied voltage and $\hat Q$ is a charge operator.

\subsection{Variational bound}
As the extended convexity holds for any ensemble of $\hat\rho$,
tighter bounds can be obtained by choosing the ensemble judiciously
\cite{alipour}. Instead of the original ensemble given by
Eq.~(\ref{ensemble}), we define a new stochastic process $Z(t)$ by
\begin{align}
X(t) = \intall d\tau g(t-\tau|\theta) Z(\tau),
\end{align}
where $g$ is an impulse-response function to be chosen later.
$\hat\rho$ can now be expressed as
\begin{align}
\hat\rho(\theta) &= \expect_{Z|\theta}
\BK{\hat U[g*Z,T]\ket\psi\bra\psi \hat U^\dagger[g*Z,T]},
\end{align}
where $*$ denotes convolution.  With
\begin{align}
\hat\sigma = \hat U[g*Z,T]\ket\psi\bra\psi \hat U^\dagger[g*Z,T],
\end{align}
this results in a family of ensembles $\{\hat\sigma,P_Z\}$
parameterized by $g$ for a given $\hat\rho$.

Assuming the Hamiltonian in Eq.~(\ref{H}), it can be shown that
\cite{twc,pasquale}
\begin{align}
J_{\mu\nu}(\hat\sigma) &= \frac{4}{\hbar^2}
\int_0^T dt \int_0^T dt' K_Q(t,t')
\nonumber\\&\quad\times
\intall d\tau \partial_\mu g(t-\tau|\theta) Z(\tau)
\nonumber\\&\quad\times
\intall d\tau' \partial_\nu g(t'-\tau'|\theta) Z(\tau'),
\label{Jdyn}
\end{align}
where
$\partial_\mu \equiv \partial/\partial\theta_\mu$ and
$K_Q(t,t')$ is the quantum covariance of the generator in the
Heisenberg picture, defined as
\begin{align}
K_Q(t,t') &\equiv 
\real\Bk{\bra\psi \Delta\hat Q(t) \Delta\hat Q(t') \ket\psi},
\label{CQ}
\\
\Delta \hat Q(t) &\equiv \hat Q(t)- \bra\psi \hat Q(t)\ket\psi,
\\
\hat Q(t) &\equiv \hat U^\dagger(X,t)\hat Q \hat U(X,t).
\end{align}
We now assume that $K_Q(t,t')$ is independent of $X(t)$; such an
assumption is commonly satisfied in linear optomechanics and
optical-phase-modulation systems.  The expected $J(\hat\sigma)$
becomes
\begin{align}
\expect_{Z|\theta}\Bk{J_{\mu\nu}(\hat\sigma)}
&= \frac{4}{\hbar^2}
\int_0^T dt \int_0^T dt' K_Q(t,t')
\nonumber\\&\quad\times
\intall d\tau \intall d\tau' K_Z(\tau,\tau'|\theta)
\nonumber\\&\quad\times
\Bk{\partial_\mu g(t-\tau|\theta)}\Bk{\partial_\nu g(t'-\tau'|\theta)},
\label{expected_qfi}
\end{align}
where
\begin{align}
K_Z(\tau,\tau'|\theta) &\equiv \expect_{Z|\theta}\Bk{Z(\tau)Z(\tau')}
\end{align}
is the prior covariance of $Z(t)$.  Assume further that the quantum
statistics of $\Delta\hat Q(t)$ are stationary, with a PSD given by
\begin{align}
S_Q(\omega) &\equiv \intall d\tau K_Q(t,t+\tau) \exp(i\omega\tau).
\end{align}
The assumption of stationary
processes and a long observation time $T$ (relative to all other time
scales in the problem) is known as the SPLOT assumption.
 Defining a transfer function as
\begin{align}
G(\omega|\theta) \equiv \intall dt g(t|\theta)\exp(i\omega t),
\end{align}
restricting $G$ to be nonzero for all frequencies of interest, noting
that the PSD of $Z(t)$ is $S_X/|G|^2$, and making the SPLOT
assumption, Eq.~(\ref{expected_qfi}) can be rewritten as
\begin{align}
\expect_{Z|\theta}\Bk{J_{\mu\nu}(\hat\sigma)}
&= T\intall \frac{d\omega}{2\pi} \frac{4S_Q S_X}{\hbar^2}
\bk{\partial_\mu\ln G}\bk{\partial_\nu\ln G^*}.
\label{Jg}
\end{align}
The Fisher information $j(P_Z)$ can be obtained by applying
Eq.~(\ref{bhatta_cfi}) to the Bhattacharyya distance between two
stationary Gaussian processes \cite{vantrees3}. The result is
\begin{align}
j_{\mu\nu}(P_Z) &= T\intall \frac{d\omega}{2\pi}
\frac{1}{2}\bk{\partial_\mu \ln \frac{S_X}{|G|^2}}
\bk{\partial_\nu \ln \frac{S_X}{|G|^2}}.
\label{jg}
\end{align}
Combining Eqs.~(\ref{Jg}) and (\ref{jg}) according to
Eq.~(\ref{extended}), we obtain
\begin{align}
u_\mu \mathcal J_{\mu\nu} u_\nu &= T\intall \frac{d\omega}{2\pi} 
\Bk{\frac{4S_Q S_X}{\hbar^2} |\lambda|^2 +
\frac{1}{2}\bk{\Lambda-\lambda-\lambda^*}^2},
\label{variational}
\nonumber\\
\lambda &\equiv u_\mu\partial_\mu \ln G,
\quad
\Lambda \equiv u_\mu\partial_\mu \ln S_X.
\end{align}
Since Eq.~(\ref{variational}) is quadratic with respect to $\lambda$,
the $\lambda$ and thus $G$ that minimizes Eq.~(\ref{variational}) for
each $u$ can be found analytically. Straightforward algebra then leads
to a variational upper bound on the quantum Fisher information given
by
\begin{align}
J &\le \tilde{\mathcal J},
&
\tilde{\mathcal J}_{\mu\nu} &\equiv 
T \intall \frac{d\omega}{2\pi}
\frac{(\partial_\mu \ln S_X)
(\partial_\nu \ln S_X)}{2+\hbar^2/(S_QS_X)}.
\label{var_bound}
\end{align}
This is the first main result of this paper. Note that the quantum
state $\ket{\psi}$ need not be Gaussian for the result to hold.

For mechanical force measurements, the straightforward choice of the
Hamiltonian leads to $S_X$ being the force PSD and $S_Q$ being the
mechanical position PSD.  For linear cavity optomechanics, the
interaction-picture purification technique explained in
Sec.~\ref{sec_dyn} leads to an alternative Hamiltonian such that $S_X$
is the PSD of the forced displacement and $S_Q$ is proportional to the
cavity photon-number PSD. For continuous optical phase modulation
\cite{wiseman_milburn,wheatley,yonezawa,iwasawa}, $S_X$ is the phase
PSD and $S_Q/\hbar^2$ is the photon-flux PSD.  In all cases, the
frequency-domain integral given by Eq.~(\ref{var_bound}), together
with the matrix inequalities
\begin{align}
\Sigma \ge j^{-1} \ge J^{-1} \ge \tilde{\mathcal J}^{-1}
\label{chain}
\end{align}
that follow from Eqs.~(\ref{crb}), (\ref{qcrb}), and
(\ref{var_bound}), represent a novel form of uncertainty relations and
indicate a nontrivial interplay between the classical noise
characterized by $S_X$ and a frequency-domain SNR given by
$S_QS_X/\hbar^2$ in bounding the estimation error and the Fisher
information quantities. Note also that $\tilde{\mathcal J}$ is
proportional to the total time $T$, as are all the Fisher information
quantities we derive here. This means that a longer observation time
can improve the parameter estimation even if the SNR is low, as is
well known in statistics \cite{shumway_stoffer} but missed by some of
the previous quantum studies \cite{nimmrichter,diosi15}.

\section{Continuous optical phase modulation}
\subsection{Error bounds}
To illustrate our theory, consider the optics experiment depicted in
Fig.~\ref{generator_process_examples}(c) or (d). An external
stochastic source $X(t)$, such as a moving mirror or an electro-optic
modulator, modulates the phase of a continuous optical beam, which is
then measured to obtain information about the source.  The Hamiltonian
is
\begin{align}
\hat H = \hbar \hat I(t)X(t),
\end{align}
where $\hat I(t)$ is the photon-flux operator, $S_X(\omega|\theta)$ is
the source PSD, and $S_I(\omega) = S_Q(\omega)/\hbar^2$ is the
photon-flux PSD. This model also applies to quantum optomechanics if
the dynamics can be linearized around a strong optical mean field and
a suitable interaction picture is used, as discussed in
Sec.~\ref{sec_dyn}. The quantum limit given by Eq.~(\ref{var_bound})
becomes
\begin{align}
\tilde{\mathcal J}_{\mu\nu} &= 
T \intall \frac{d\omega}{2\pi}
\frac{(\partial_\mu \ln S_X)(\partial_\nu \ln S_X)}
{2+1/(S_IS_X)}.
\label{Jphase}
\end{align}
Equation~(\ref{Jphase}) together with Eq.~(\ref{chain}) represent an
uncertainty relation between the phase spectrum-parameter estimation
error and the photon-flux PSD.

\begin{figure}[htbp!]
\centerline{\includegraphics[width=0.45\textwidth]{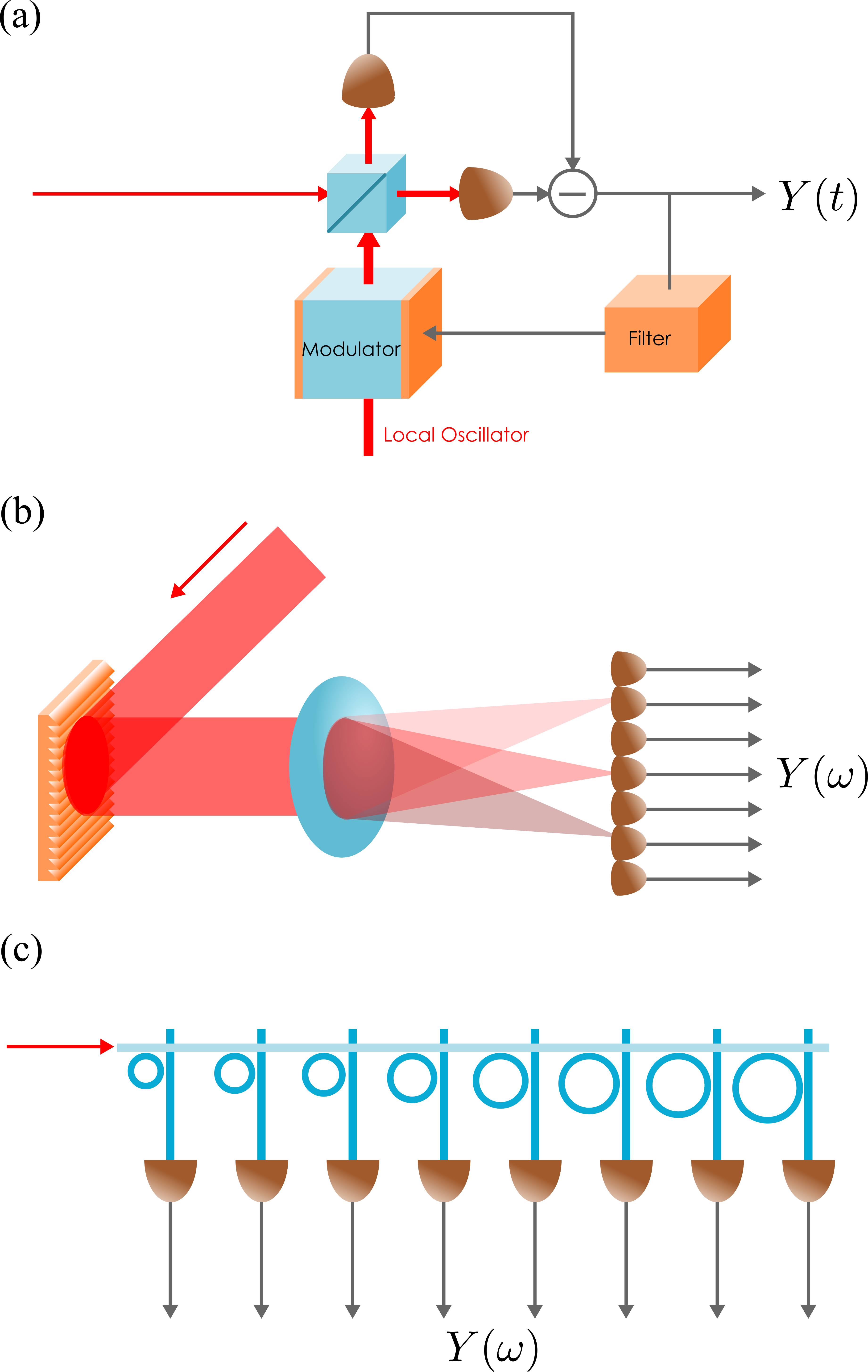}}
\caption{(Color online). (a) Adaptive homodyne detection. (b) Spectral
  photon counting with a diffraction grating and a lens. (c) Spectral
  photon counting with an optical-resonator array.}
\label{measurements}
\end{figure}

We can compare our bound with the Fisher information for
homodyne detection, a standard experimental phase measurement method
\cite{wiseman_milburn,wheatley,yonezawa,iwasawa}, as illustrated in
Fig.~\ref{measurements}(a).  If the mean field is strong, and the
modulation is weak or tight phase locking is achieved, the output
process can be linearized as 
\begin{align}
Y(t) \approx X(t) + \eta(t),
\end{align}
where $\eta(t)$ is the phase-quadrature noise.  The information
$j(P_Y^{(\rm hom)})$ can be computed analytically if $\eta$ is
Gaussian and stationary with power spectral density $S_\eta(\omega)$
such that $Y$ is also Gaussian and stationary \cite{vantrees3}; the
result with the SPLOT assumption is
\begin{align}
j_{\mu\nu}\bk{P_Y^{(\rm hom)}} 
= T \intall \frac{d\omega}{2\pi}
\frac{(\partial_\mu \ln S_X)(\partial_\nu \ln S_X)}
{2(1 + S_\eta/S_X)^2}.
\end{align}
The classical Cram\'er-Rao bound $\Sigma \ge j^{-1}(P_Y^{(\rm hom)})$
is asymptotically attainable for long $T$ using maximum-likelihood
estimation \cite{shumway_stoffer}.

With the quadrature uncertainty relation
\begin{align}
S_\eta(\omega)S_I(\omega) \ge \frac{1}{4}
\end{align}
for the optical beam \cite{walls_milburn}, the optimal homodyne
information is
\begin{align}
j\bk{P_Y^{(\rm hom)}} &\le \tilde j,
\nonumber\\
\tilde j_{\mu\nu}
&\equiv T \intall \frac{d\omega}{2\pi}
\frac{(\partial_\mu \ln S_X)(\partial_\nu \ln S_X)}
{2 + 1/(S_IS_X) + 1/(8S_I^2S_X^2)}.
\label{jhom}
\end{align}
We can compare this homodyne limit with the quantum limit in
Eq.~(\ref{Jphase}); the expressions are similar, apart from a extra
factor of $1/(8S_I^2S_X^2)$ that makes the homodyne limit strictly
worse than our quantum limit, especially if $S_IS_X$ is small.

\subsection{Spectral photon counting}
Although Eq.~(\ref{chain}) sets rigorous lower bounds on the
estimation error $\Sigma$, there is no guarantee that the the error
for any measurement can attain the final bound $\tilde J^{-1}$.
Inspired by our previous work on astronomical quantum optics
\cite{stellar,nair_tsang}, here we analyze an alternative measurement
that we call spectral photon counting. Physically, it is simply a
conventional optical spectrometer with photon counting for each
spectral mode \cite{shapiro98,brady}. The first step of spectral
photon counting is the coherent optical Fourier transform via a
dispersive optical element, such as a diffraction grating or a prism
and a Fourier-transform lens \cite{shapiro98} as depicted in
Fig.~\ref{measurements}(b), or an array of optical ring resonators
with different resonant frequencies coupled to a cross grid of
waveguides \cite{chu99} as depicted in Fig.~\ref{measurements}(c). The
second step is a measurement of the photon numbers in the spectral
modes, and the final step is a maximum-likelihood estimation of
$\theta$ from the spectral photon counting results. For the phase
spectrum-parameter estimation problem with weak modulation and a
coherent-state input, this method turns out to have an information
$j(P_Y^{(\rm spc)})$ coinciding with $\tilde{\mathcal J}$ for all
parameters.

Let the positive-frequency electric field at the input of the phase
modulator be 
\begin{align}
\hat E^{(+)}(t) = \hat A(t)\exp(-i\Omega t),
\end{align}
where $\hat A(t)$ is an annihilation operator for the slowly varying
envelope with commutation relation
\begin{align}
[\hat A(t),\hat A^\dagger(t')] = \delta(t-t'),
\end{align}
and $\Omega$ is the optical carrier frequency. With a strong mean
field 
\begin{align}
\alpha \equiv \bra\psi\hat A(t)\ket\psi
\end{align}
and weak phase modulation, the output field can be linearized as
\begin{align}
\hat B(t) \approx \hat A(t) + i\alpha X(t).
\end{align}
To model the optical Fourier transform, we follow Shapiro
\cite{shapiro98} to express each frequency mode in terms of the mode
annihilation operator as
\begin{align}
\hat b_m = \frac{1}{\sqrt{T}}\int_0^{T} dt \hat B(t)\exp(i\omega_m t),
\end{align}
with sideband frequencies 
\begin{align}
\omega_m &= \frac{2\pi m}{T},
&
m &\in \BK{\dots,-2,-1,0,1,2,\dots},
\end{align}
and
\begin{align}
[\hat b_m,\hat b_n^\dagger] = \delta_{mn}.
\end{align}
Assuming $\alpha$ to be time-constant,
\begin{align}
\hat b_m \approx \hat a_m + i\alpha x_m,
\end{align}
where $\hat a_m$ is the Fourier transform of $\hat A(t)$ and $x_m$ is
that of $X(t)$ in the same way as $\hat b_m$.

The strong mean field is contained in the $m = 0$ mode only, and if
the spectrum of $x_m$ is wide, negligible information is lost if we
neglect the $m = 0$ mode. The other modes are coherent states for a
given displacement $i\alpha x_m$ if the input beam is a coherent state
\cite{shapiro98}. For a given $x_m$, the photon-counting distribution
for $\hat n_m \equiv \hat b_m^\dagger \hat b_m$ in each mode is
therefore Poissonian with mean $|\alpha|^2|x_m|^2$ and
independent from one another.

Since $X(t)$ is a hidden stochastic process, we must average the
Poissonian distribution over the prior of $X(t)$ to obtain the final
likelihood function. For a Gaussian $X(t)$ with the SPLOT assumption,
$\{x_m; m > 0\}$ are independent complex Gaussian random variables
with variances $S_X(\omega_m|\theta)$ \cite{shumway_stoffer}, but
since $X(t)$ is real, the sidebands are symmetric with
$x_m = x_{-m}^*$. This means that, averaged over $x$, the
photon numbers at opposite sideband frequencies become correlated.

To simplify the analysis, suppose that, for each $m > 0$, we sum the
pair of measured photon numbers $n_m$ and $n_{-m}$ at opposite
sidebands and use a reduced set of measurement record
$\{N_m \equiv n_m + n_{-m}; m >0\}$ for estimation. It can be shown
that each $N_m$ is also Poissonian conditioned on the mean
$2|\alpha|^2|x_m|^2$, but now they remain independent from one another
in the set after averaging over $\{x_m; m> 0\}$.

With $x_m$ being complex Gaussian and $N_m$ being conditionally
Poissonian with mean $2|\alpha|^2|x_m|^2$, it can be shown that the
marginal distribution of $N_m$ is a Bose-Einstein distribution
\cite{walls_milburn} with mean number
\begin{align}
\bar N_m = 2|\alpha|^2 S_X(\omega_m|\theta).
\end{align}
The Fisher information remains analytically
tractable and is given by
\begin{align}
j_{\mu\nu}(P_Y^{(\rm spc)}) = \sum_{m > 0} \frac{(\partial_\mu \ln \bar
N_m) (\partial_\nu \ln \bar N_m)}{1+1/\bar N_m}.
\end{align}
If we use the SPLOT assumption to
replace $\sum_{m > 0}$ with $T\int_0^\infty d\omega/(2\pi)$
\cite{helstrom} and use the symmetry of the integrand to replace
$T\int_0^\infty d\omega/(2\pi)$ with $(T/2)\intall d\omega/(2\pi)$,
 the Fisher information becomes
\begin{align}
j_{\mu\nu}\bk{P_Y^{(\rm spc)}} &= T\intall \frac{d\omega}{2\pi} 
\frac{(\partial_\mu \ln S_X)(\partial_\nu \ln S_X)}
{2+1/(\mathcal N S_X)},
\label{jspc}
\end{align}
where $\mathcal N$ is the average input photon flux.  Since
$S_I(\omega) = \mathcal N$ for a coherent state, Eq.~(\ref{jspc})
coincides with the quantum bound in Eq.~(\ref{Jphase}).  This is the
second main result of this paper. Comparing Eq.~(\ref{jspc}) with the
homodyne limit given by Eq.~(\ref{jhom}), we can expect that spectral
photon counting becomes significantly better than homodyne detection
when $\mathcal NS_X$ is small.

\subsection{Ornstein-Uhlenbeck spectrum analysis}
For a more specific example, consider the experiments in
Refs.~\cite{wheatley,yonezawa}, which can be modeled as the
continuous-optical-phase-modulation problem depicted in
Fig.~\ref{generator_process_examples}(d), with adaptive homodyne
detection depicted in Fig.~\ref{measurements}(a) and $X(t)$ given by
an Ornstein-Uhlenbeck process.  The PSD of $X(t)$ is
\begin{align}
S_X(\omega|\theta) &= \frac{2\theta_1\theta_2}{\omega^2+\theta_2^2},
\label{ou}
\end{align}
where $\theta_1 = \expect_{X|\theta}[X^2(t)]$ is the area under $S_X$
and $\theta_2$ is the bandwidth. The experimental $S_I$ can be assumed
to be constant for all frequencies of interest, and the quantum limit
given by Eq.~(\ref{Jphase}) on the estimation of $\theta_1$ and
$\theta_2$ can be computed analytically:
\begin{align}
\tilde{\mathcal J}_{11} &= \frac{\theta_2 T}{8\theta_1^2} \frac{C}{\sqrt{1+C/2}},
\nonumber\\
\tilde{\mathcal J}_{22} &= \frac{2T}{\theta_2}
\frac{1+C/4}{C}\bk{\frac{1+C/4}{\sqrt{1+C/2}}-1},
\nonumber\\
\tilde{\mathcal J}_{12} = \tilde{\mathcal J}_{21} &= \frac{T}{2\theta_1}
\bk{\frac{1+C/4}{\sqrt{1+C/2}}-1},
\label{J_ou}
\end{align}
where 
\begin{align}
C & \equiv \frac{8\theta_1 S_I}{\theta_2} = 4S_I S_X(0|\theta)
\end{align}
is an SNR quantity. For comparison, the homodyne limit given by
Eq.~(\ref{jhom}) is
\begin{align}
\tilde j_{11} &= 
\frac{\theta_2 T}{8\theta_1^2} \frac{C^2}{(1+C)^{3/2}},
\nonumber\\
\tilde j_{22} &=
\frac{2T}{\theta_2}\frac{1}{C}
\left[\frac{(1+C/2)(1+5C/4+C^2/8)}{(1+C)^{3/2}}\right.
\nonumber\\
&\quad \left.-\bk{1+\frac{C}{4}}\right],
\nonumber\\
\tilde j_{12} = \tilde j_{21} &= 
\frac{T}{2\theta_1}
\Bk{\frac{1+3C/2+C^2/4}{(1+C)^{3/2}}-1}.
\label{jhom_ou}
\end{align}
For homodyne detection, $C$ is an upper limit on the ratio between the
peak of $S_X$ and the homodyne noise floor $S_\eta$ in the frequency
domain.

\begin{figure}[htbp!]
\centerline{\includegraphics[width=0.45\textwidth]{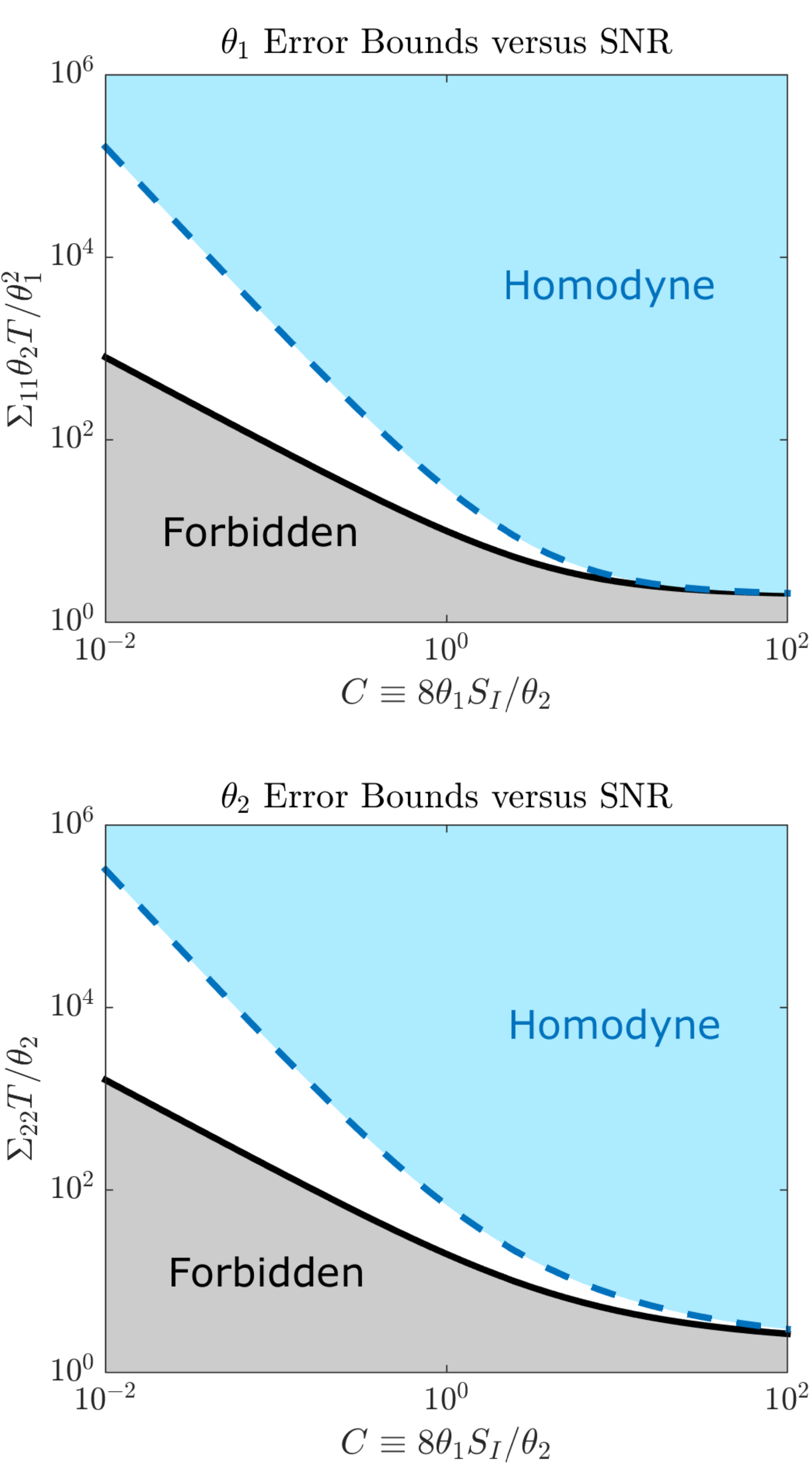}}
\caption{(Color online). Log-log plots of the quantum limit
  $\tilde{\mathcal J}^{-1}$ (inverse of Eqs.~(\ref{J_ou}), black solid
  line) and homodyne limit $\tilde j^{-1}$ (inverse of
  Eqs.~(\ref{jhom_ou}), blue dashed line) on the mean-square errors
  versus an SNR quantity $C \equiv 8\theta_1 S_I/\theta_2$.  Top plot:
  limits on $\Sigma_{11}$ (normalized in a unit of
  $\theta_1^2/(\theta_2 T)$), bottom plot: limits on $\Sigma_{22}$
  (normalized in a unit of $\theta_2/T$). No measurement can achieve
  an error below the quantum limit (grey ``forbidden'' region), while
  the homodyne performance (blue ``homodyne'' region) cannot go below
  the homodyne limit. For $C \gg 1$, the limits approach constants,
  while for $C \ll 1$ the homodyne limit has a significantly worse
  error scaling.}
\label{bounds}
\end{figure}

Figure~\ref{bounds} plots the quantum ($\tilde{\mathcal J}^{-1}$) and
homodyne ($\tilde j^{-1}$) bounds on the estimation errors
$\Sigma_{11}$ and $\Sigma_{22}$ versus $C$.  Both plots show similar
behaviors, and the $C \gg 1$ and $C \ll 1$ limits are of special
interest. In the high-SNR regime ($C \gg 1$), both
$\tilde{\mathcal J}^{-1}$ and $\tilde j^{-1}$ approach a
$C$-independent limit:
\begin{align}
\lim_{C\to \infty}\tilde{\mathcal J}^{-1} &
=\lim_{C\to\infty} \tilde j^{-1} = \frac{2}{\theta_2 T}
\bk{\begin{array}{cc}\theta_1^2 & -\theta_1\theta_2\\
-\theta_1\theta_2  & \theta_2^2\end{array}},
\label{highC}
\end{align}
and the homodyne performance is near-quantum-optimal.  This asymptotic
behavior is different from that of the bounds for single-parameter
estimation, as both $1/\tilde{\mathcal J}_{\mu\mu}$ and
$1/\tilde j_{\mu\mu}$ scale as $C^{-1/2}$ and decrease indefinitely
for increasing $C$. The matrix bounds thus demonstrate the detrimental
effect of having two unknown parameters that act as noise to each
other. The $C$-independent limits also suggest that, once an
experiment is in the high-SNR regime, no significant improvement can
be made by increasing $S_I$ and reducing the noise floor via
photon-flux increase, squeezing, or changing the measurement method.

In the low-SNR regime ($C \ll 1$), on the other hand, it can be shown
that
\begin{align}
\tilde{\mathcal J}^{-1} &\approx \frac{8}{\theta_2 T} C^{-1}
\bk{\begin{array}{cc}\theta_1^2 & 0\\
0  & 2\theta_2^2\end{array}},
\label{lowCQ}
\\
\tilde j^{-1} &\approx \frac{16}{\theta_2 T}C^{-2}
\bk{\begin{array}{cc}\theta_1^2 & \theta_1\theta_2\\
\theta_1\theta_2  & 2\theta_2^2\end{array}},
\label{lowCH}
\end{align}
where the homodyne bounds on $\Sigma_{11}$ and $\Sigma_{22}$ diverge
from the quantum bounds by a large factor of $2/C \gg 1$. The
diverging bounds demonstrate the importance of quantum-optimal
measurement in the low-SNR limit: at least for a coherent-state input
and weak modulation, the quantum-optimal performance of spectral
photon counting can exhibit a superior error scaling and offer
significant improvements over homodyne detection.

\subsection{Experimental data analysis}
To compare our theory with actual experimental performance, we analyze
the data from the experiment reported in Ref.~\cite{wheatley}, which
is in a high-SNR regime ($C \ge 23.5$) and the adaptive homodyne
performance is expected to be close to our quantum limit. We focus on
the experiment with coherent states and not the one with squeezed
states reported in Ref.~\cite{yonezawa}, as Eqs.~(\ref{highC}) imply
that squeezing offers insignificant improvement in this high-SNR
regime. 

The experiment reported in Ref.~\cite{wheatley} used four different
mean photon fluxes $\mathcal N_1 = 1.315\times 10^6~{\rm s}^{-1}$,
$\mathcal N_2 = 3.616\times 10^6~{\rm s}^{-1}$,
$\mathcal N_3 = 6.327\times 10^6~{\rm s}^{-1}$,
$\mathcal N_4 = 1.418\times 10^7~{\rm s}^{-1}$.  For each photon flux
$\mathcal N_k$, $M_k$ traces of $X(t)$ and $M_k$ traces of $Y(t)$ were
recorded ($M_1 = 21$, $M_2 = 23$, $M_3 = 24$, $M_4 = 27$). Each trace
of $Y(t)$ was obtained using a different feedback gain for the filter
in the phase-locked loop, such that the phase locking might not be
optimal. The original purpose of varying the feedback gains was to
demonstrate the existence of an optimal filter for phase estimation in
Ref.~\cite{wheatley}, but it is also coincidentally appropriate in our
present context, as $\theta_1$ and $\theta_2$ are supposed to be
unknown here and the optimal filter is not supposed to be known. To
make the data analysis tractable, we assume that the phase locking
remained tight even if the filter was suboptimal, such that we can
still use the linearized model
\begin{align}
Y(t) &= \sin[X(t)-\check X(t)] +\eta(t) + \check X(t) 
\approx X(t) +\eta(t),
\label{linear_Y}
\end{align}
where $\check X(t)$ is the feedback phase modulation on the local
oscillator. Comparisons of the experimental $X(t)$ with $\check X(t)$
show that $\expect[X(t)-\check X(t)]^2 \lesssim 0.3$ and the
linearized model is reasonable. Most metrological experiments, such as
gravitational-wave detectors, deal with extremely weak phase
modulation, so the linearized model is expected to be even more
accurate in those cases. Appendix~\ref{sec_cal} describes further
calibrations to ensure that Eq.~(\ref{linear_Y}) is accurate.

For any observation time $T$, the maximum-likelihood estimation can be
performed using an expectation-maximization algorithm
\cite{shumway_stoffer,ang}, but our numerical simulations suggest that
it is safe here to use a simpler and faster method due to Whittle
\cite{whittle}, which exploits the SPLOT assumption to simplify the
likelihood function. Consider a real discrete-time series
\begin{align}
\BK{Y(t_l);l = 0,1,\dots,L-1},  
\quad t_l = l\delta t, 
\end{align}
and zero-mean
Gaussian statistics conditioned on $\theta$. Define the discrete
Fourier transform as
\begin{align}
y_m &= \frac{\delta t}{\sqrt{T}} \sum_{l=0}^{L-1} Y(t_l)\exp(i \omega_m t_l),
&
\omega_m &= \frac{2\pi m}{T},
\end{align}
with integer $m$ and $y_m = y_{L-m}^*$. It can be shown that, with the
SPLOT assumption, the positive-frequency components
$\{y_m; 0 < m < L/2\}$ are independent zero-mean complex Gaussian
random variables with variances $S_Y(\omega_m|\theta)$
\cite{whittle,shumway_stoffer}. This means that the log-likelihood
function, up to a $\theta$-independent additive constant $\mathcal A$,
can be approximated as
\begin{align}
\ln P_Y &\approx \mathcal A -\sum_{0< m < L/2}
\Bk{\ln S_Y(\omega_m|\theta) + \frac{|y_m|^2}{S_Y(\omega_m|\theta)}}.
\label{whittle}
\end{align}
Approximate maximum-likelihood estimation can then be performed by
Fourier-transforming the time series into $\{y_m\}$ and finding the
parameters that maximize Eq.~(\ref{whittle}).  We use Matlab\circledR\
and its \texttt{fft} and \texttt{fminunc} functions to implement this
procedure on a desktop PC. With $T = 0.01$~s for each $Y(t)$ trace, we
expect the SPLOT assumption to be reasonable. We also perform
numerical simulations throughout our analysis to ensure that our SPLOT
and unbiased-estimator assumptions are valid and our results are
expected.

To prevent technical noise and model mismatch at higher frequencies
from contaminating our analysis, we consider only the spectral
components up to $6\times 10^5$~rad/s $\sim 10\theta_2$, rather than
the full measurement bandwidth $\pi/\delta t = \pi\times 10^8$~rad/s.
To estimate the true parameters more accurately, we apply the Whittle
method to the collective record of all $\sum_k M_k = 95$ experimental
$X(t)$ traces, assuming the spectrum given by Eq.~(\ref{ou}), and
obtain $\theta_1 = 0.1323$ and $\theta_2 = 5.909\times 10^4$~rad/s. We
take these to be the true parameters, as the estimates from such a
large number of $X(t)$ traces are expected to be much more accurate
than those from each $Y(t)$ trace.

\begin{figure}[htbp!]
\centerline{\includegraphics[width=0.45\textwidth]{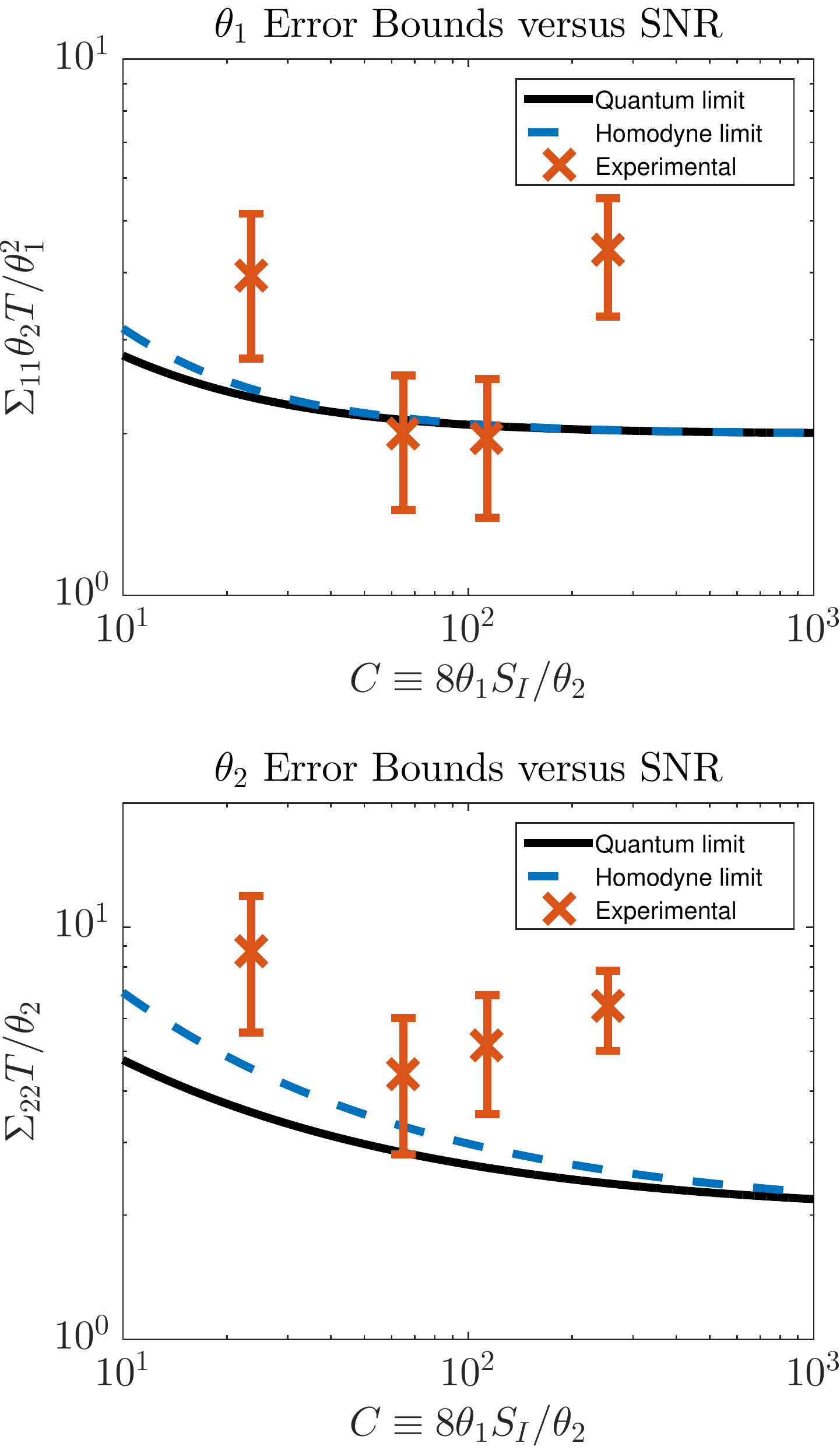}}
\caption{(Color online). Log-log plots of the quantum limit
  $\tilde{\mathcal J}^{-1}$ (inverse of Eqs.~(\ref{J_ou}), black solid
  line), the homodyne limit $\tilde j^{-1}$ (inverse of
  Eqs.~(\ref{jhom_ou}), blue dash line), and the experimental
  mean-square estimation errors $\Sigma$ versus the SNR quantity
  $C \equiv 8\theta_1 S_I/\theta_2$.  Top plot: Experimental
  $\Sigma_{11}=\{4.0\pm 1.2, 2.0 \pm 0.6, 2.0 \pm 0.6, 4.4\pm 1.1\}$
  (in a unit of $\theta_1^2/(\theta_2 T)$) versus
  $C = \{23.5, 64.8, 113, 254\}$, compared with the homodyne limit and
  the quantum limit. Bottom plot: Experimental
  $\Sigma_{22} = \{8.7\pm 3.2, 4.4\pm 1.6, 5.2\pm 1.7, 6.4\pm 1.4\}$
  (in a unit of $\theta_2/T$) versus the same $C$ values, compared
  with the homodyne limit and the quantum limit.}
\label{plots}
\end{figure}

We apply the Whittle method to each $Y(t)$ trace and
evaluate the estimation errors by comparing the estimates with the
true parameters. For each photon flux we assume a noise floor that is
estimated from high-frequency data, and then we estimate $\theta$
using spectral components of $Y$ up to $\omega = 6\times 10^5$~rad/s.
Let the resulting estimates be
\begin{align}
\BK{\check\theta_{\mu k}^{(m_k)};\mu = 1,2;k = 1,2,3,4; m_k =
1,\dots,M_k},
\end{align}
where $\mu$ is the index for the two parameters, $k$ is the index for
the photon fluxes, and $m_k$ is the index for the traces, and let the
squared distance of each estimate from the true parameter be
\begin{align}
\varepsilon_{\mu k}^{(m_k)} \equiv \bk{\check\theta_{\mu
  k}^{(m_k)}-\theta_\mu}^2.
\end{align}
$\varepsilon_{\mu k}^{(m_k)}$ can be regarded as an outcome for a
random variable $\varepsilon_{\mu k}$, so we can use the sample mean
\begin{align}
\bar \varepsilon_{\mu k} \equiv \frac{1}{M_k}\sum_{m_k=1}^{M_k}
\varepsilon_{\mu k}^{(m_k)}
\end{align}
to estimate the expected error
\begin{align}
\Sigma_{\mu\mu} = \expect_Y(\varepsilon_{\mu k}).
\end{align}
To find the deviation of the sample mean $\bar \varepsilon_{\mu k}$
from the expected value, we use an unbiased estimate of the variance
of $\varepsilon_{\mu k}$, that is,
\begin{align}
V_{\mu k} \equiv \frac{1}{M_k-1}\sum_{m_k=1}^{M_k} 
\bk{\varepsilon_{\mu k}^{(m_k)}-\bar\varepsilon_{\mu k}}^2,
\end{align}
and divide it by the number of samples $M_k$.  Our final results
\begin{align}
  \BK{\bar \varepsilon_{\mu k} \pm \sqrt{\frac{V_{\mu k}}{M_k}};
\mu = 1,2; k =  1,2,3,4}
\end{align}
are plotted in normalized units in Fig.~\ref{plots}, together with the
quantum limit given by the inverse of Eqs.~(\ref{J_ou}) and the
homodyne limit given by the inverse of Eqs.~(\ref{jhom_ou}).  The
plots demonstrate estimation errors close to both the homodyne limit
and the fundamental quantum limit, despite experimental imperfections
such as imperfect phase locking.

\section{Conclusion}
We have presented three key results in this paper: a
measurement-independent quantum limit to spectrum-parameter
estimation, the optimality of spectral photon counting, and an
experimental data analysis. The quantum limit applies to a wide range
of experiments and is particularly relevant to optomechanics, where
the spectrum parameters of a stochastic force are often of interest to
gravitational-wave astronomy
\cite{ligo09,nimmrichter,diosi15,ligo16}. The proposed spectral photon
counting method will be useful whenever the problem can be modeled as
weak phase modulation of a coherent state and the SNR is low. Most
metrological experiments, including gravitational-wave detectors,
involve extremely weak phase modulation and low SNR, so the potential
improvement over homodyne or heterodyne detection without the need of
squeezed light is an important discovery.  Our experimental data
analysis further demonstrates the relevance of our theory to current
technology and provides a recipe for future spectrum-analysis
experiments.

There are many interesting potential extensions of our
theory. Although quantum baths can often be modeled classically, a
generalization of our formalism to account explicitly for nonclassical
baths will make our theory applicable to an even wider range of
experiments. A generalization for nonstationary processes and finite
observation time will be valuable for the study of unstable systems,
which are potentially more sensitive than stable systems
\cite{loschmidt}.  Tighter quantum limits that explicitly account for
decoherence may be derived by applying the techniques in
Refs.~\cite{escher,demkowicz,tsang_open}.  A Bayesian formulation that
removes the unbiased-estimator assumption should be possible
\cite{vantrees,twc,tsang_open,qzzb,qbzzb}. A more detailed study of
our theory in the context of optomechanics can serve as an extension
of Refs.~\cite{nimmrichter,diosi15} and enable a more rigorous
analysis of quantum limits to testing wavefunction-collapse
models. Application of our theory to spin systems will provide a more
rigorous foundation for stochastic magnetometry \cite{hall09}.

The actual performance of spectral photon counting depends on the
bandwidth and spectral resolution of the Fourier-transform device, as
well as the quantum efficiency and dark counts of the photodetectors
in practice. While a more detailed analysis of such practical concerns
is needed before one can judge the realistic performance of spectral
photon counting with current technology, the large potential
improvement in the low-SNR regime indicates the fundamental importance
of coherent optical information processing for sensing applications
and should motivate further technological advances in coherent quantum
optical devices
\cite{carolan,stellar,gottesman,nair_tsang,tnl,sliver,tnl2}. In the
high-SNR regime, on the other hand, our theory and experimental data
analysis suggest that current technology can already approach the
quantum limits with homodyne or even heterodyne detection. In this
regime, our quantum limit primarily serves as a no-go theorem, proving
that no other measurement can offer significant improvement. The
challenge for actual metrological experiments will be to reach the
high-SNR regime for weak signals, in which case our theory should
serve as a rigorous foundation to guide future experimental designs.

\section*{Acknowledgments}
We acknowledge helpful discussions with Ranjith Nair and Xiao-Ming Lu.
This work is supported in part by the Singapore National Research
Foundation under NRF Grant No.~NRF-NRFF2011-07, Singapore Ministry of
Education Academic Research Fund Tier 1 Project R-263-000-C06-112, the
Australian Research Council, Grant No.~CE1101027, PDIS, GIA, APSA
commissioned by the MEXT of Japan, ASCR-JSPS and the SCOPE program of
the MIC of Japan, CREST of JST.

\appendix
\section{\label{sec_cal}Experimental data recalibration}
In the experiment described in Ref.~\cite{wheatley}, calibration
procedures were used to convert applied and measured voltages to the
various physical quantities defined throughout Ref.~\cite{wheatley}.
In the course of analysing that experimental data for the purposes of
the new estimation task described here, we found that the data gives
non-negligible bias in the estimation of $\theta_1$. It turns out that
the original calibration of experimental data was not accurate enough
for the new task of estimating $\theta_1$ (note that $\theta_2$ is
robust against this inaccuracy).  The systematic calibration error had
insignificant effects on the phase estimation task in
Ref.~\cite{wheatley} -- making the estimate slightly worse than it
would have been without the bias but generally within the uncertainty
of the experiment as reported in Ref.~\cite{wheatley}. The bias might
have been caused by non-linearity or saturation of electronic circuits
during the calibration phase of the experiment or long timescale
drift.  For the purpose of this new estimation task, we refine the
calibration of the data from Ref.~\cite{wheatley} so that we can
achieve an accurate estimate. To do this in a fair way we use two
extra data sets ($k=5,6$), which were not shown in
Ref.~\cite{wheatley} but recorded by the same experimental setup with
different experimental parameters. 
Mean photon fluxes of these data sets are
$\mathcal N_5 = 6.198\times 10^6~{\rm s}^{-1}$ and
$\mathcal N_6 = 5.986\times 10^6~{\rm s}^{-1}$.  Number of traces are
$M_5=24$ and $M_6=24$.
Note that we use these ``training'' data only for the purposes of
refining the experimental calibration.  We apply the Whittle method to
the two extra data sets to obtain the true $\theta_{1}$ from the
collective record of $X(t)$, and a mean value of the estimated
$\theta_{1}$ from the collective record of $Y(t)$ traces using the
coarse calibration from Ref.~\cite{wheatley}. We determine that a
refined calibration factor of 0.8945 is required to cancel the
unwanted bias in the estimate of $\theta_{1}$ for the extra data sets
$k=5,6$. We then apply the refined calibration factor to $Y(t)$ of the
original data sets ($k=1$ to 4). By this method, we can refine the
calibration of the original data presented in Ref.~\cite{wheatley} by
making use of independent, but contemporaneously recorded data.

\bibliography{research}

\end{document}